\documentclass[aps,prl,twocolumn,showpacs,superscriptaddress]{revtex4-1}

\usepackage[]{graphicx}
\usepackage[colorlinks=true,citecolor=blue]{hyperref} 

\begin{document}


\title{High Q mg-scale monolithic pendulum for quantum-limited gravity measurements}


\author{Seth B. Cata\~{n}o-Lopez}
\email[]{seth@quantum.riec.tohoku.ac.jp}
\author{Jordy G. Santiago-Condori}
\author{Keiichi Edamatsu}
\affiliation{Research Institute of Electrical Communication, Tohoku University, Sendai 980-8577, Japan}
\author{Nobuyuki Matsumoto}
\email[]{matsumoto.granite@gmail.com}
\affiliation{Research Institute of Electrical Communication, Tohoku University, Sendai 980-8577, Japan}
\affiliation{Frontier Research Institute for Interdisciplinary Sciences, Tohoku University, Sendai 980-8578, Japan}
\affiliation{JST, PRESTO, Kawaguchi, Saitama 332-0012, Japan}


\date{\today}

\begin{abstract}
We present the development of a high $Q$ monolithic silica pendulum weighing 7 mg. The measured $Q$ value for the pendulum mode at 2.2 Hz was  $2.0 \times 10^6$. To the best of our knowledge this is the lowest dissipative mg-scale mechanical oscillator to date. By employing this suspension system, the optomechanical displacement sensor for gravity measurements we recently reported in   \href{https://doi.org/10.1103/PhysRevLett.122.071101}{Phys. Rev. Lett. 122, 071101 (2019)} can be improved to realize quantum-noise-limited sensing at several hundred Hz. In combination with the optical spring effect, the amount of intrinsic dissipation measured in the pendulum mode is enough to satisfy requirements for measurement-based quantum control of a massive pendulum confined in an optical potential. 
This paves the way for not only testing dark matter via quantum-limited force sensors, but also Newtonian interaction in quantum regimes, namely, between two mg-scale oscillators in quantum states, as well as improving the sensitivity of gravitational-wave detectors.
\end{abstract}


\maketitle

{\it Introduction.}---
The development of quantum-limited displacement sensors of macroscopic mechanical oscillators is a key component for the direct measurement and investigation of macroscopic quantum mechanics \cite{Chen2013}, the quantum nature of Newtonian interaction \cite{Miao2019,Kafri2014}, direct detection of dark matter by looking at “fifth forces”  \cite{Graham2016,Carney2019}, continuous spontaneous localization (CSL) models \cite{Bassi2013}, and gravitational-wave (GW) astronomy \cite{Abbott2016}.
Partly motivated by this, cavity optomechanics has pioneered the development of low-loss mechanical oscillators in a variety of different architectures \cite{Aspelmeyer2014,Tsaturyan2017,Ghadimi2018}, opening the door to measurement-based control of mechanical oscillators in the quantum regime \cite{Rossi2018,Cripe2019,Sudhir2017}. On the other hand, recent proposals to investigate gravitational interactions at the mg scale \cite{Schmole2016} have motivated the top-down approach relying on techniques utilizing macroscopic suspended pendulums, getting inspiration from GW detectors. However, the development of a mechanical oscillator with the possibility of quantum-limited sensitivity, while at the same time being massive enough to measure gravitational interactions has yet to be realized.
 
To achieve measurement-based quantum control, the oscillator must satisfy two basic requirements \cite{Braginskii1969,Braginsky1992,Braginsky1999}. The first demands for the frequency of oscillation to exceed the thermal decoherence rate, which induces heating from the thermal bath into the system i.e., $\omega_m>\bar{n}\gamma_m$. Here $\bar{n}$ is the average phonon number of the oscillating mode, $\omega_m/2\pi$ is its resonance frequency, and $\gamma_m$ is the oscillating mode's dissipation. This translates into the requirement for the commonly named “$Qf$ product” \cite{Braginsky1992,Aspelmeyer2014}: 
\begin{equation}
Q_m\omega_m>k_BT/\hbar,
\label{fq}
\end{equation}
which establishes a lower bound on the quality factor $Q_m=\omega_m/\gamma_m$ of the mode, necessary to undergo at least one coherent mechanical oscillation before one phonon from the thermal bath enters the mode. Here $k_B$ is the Boltzmann constant, $\hbar$ is the reduced Planck constant, and $T$ is the temperature of the thermal bath. 

The second requirement is closely related to the minimum readout noise required to resolve the zero-point motion of the oscillator $x_{\mathrm{zpf}}=\sqrt{\hbar/2m\omega_m}$ in a measurement time scale faster than the thermal decoherence rate. In optomechanical systems using massive pendulums, we can set the standard quantum noise limit for a free mass $S_{\mathrm{SQL}}=\sqrt{2\hbar/m\omega^2}$ (hereinafter “SQL”) \cite{Braginsky1968,Braginsky1992,Chen2013} as the reference readout noise level. This is because the sum of the main readout noises, like shot noise and mirror thermal noise, can be designed to be close to the SQL at several hundred Hz \cite{Matsumoto2014,Martynov2016,Aso2013}. This translates into our noise requirement in terms of dissipation: 
\begin{equation}
\frac{\omega^2}{\gamma_m}>4k_BT/\hbar .
\label{mesurementrate}
\end{equation}

In these expressions $\omega$ is the Fourier frequency. This second requirement is critical for the oscillator to be implemented in any type of measurement-based quantum experiment like feedback cooling.
In optomechanical experiments implementing pendulums, an optical spring can be used to trap and shift the pendulum mode to higher frequencies \cite{Braginskii1967,Khalili2001}. This effect does not add excess thermal fluctuating forces on the pendulum, since even 
at room temperature the optical field has a thermal occupation of almost zero. Thus, when the second condition is satisfied at some frequency, the first condition can also be satisfied by changing the pendulum's frequency around that frequency band. In terms of optomechanical parameters, this means the quantum cooperativity $C_q$ is over 1 in that range of frequencies (the cooperativity is a parameter comparing the coupling strength of the system and its dissipation; see \cite{Aspelmeyer2014} for a more detailed discussion). In other words, the condition $C_q>1$ describes a situation in which the quantum back-action of the measurement exceeds the effect of the thermal Brownian motion. 

In this Letter, we present the development of a monolithic mg-scale silica pendulum with an intrinsic pendulum quality factor of $Q_m=2.0\times 10^6$ at 2.2 Hz, capable of satisfying both requirements between 400-1800 Hz. Implementing an optical spring in this frequency range is within the capability of previously reported experiments \cite{Matsumoto2019,Matsumoto2016a,Corbitt2007a}, and therefore paves the way to the study of a mg-scale oscillator's motion in the quantum regime, and test of the intersection between gravitational and quantum regimes.

{\it Pendulum as system.}---
Under the fluctuation-dissipation theorem, interaction with the environment produces a fluctuating force on a mechanical oscillator dependent on its dissipation \cite{Callen1951}. A pendulum system by itself can allow for the pendulum mode's $Q$ value to exceed by orders of magnitude the upper bound imposed on it by intrinsic material dissipation. Therefore, massive oscillators have traditionally been isolated via suspension pendulums to achieve minimal dissipation. This effect is termed dissipation dilution because the energy loss is being diluted by the intrinsically lossless gravitational potential, where most of the energy is stored. The ratio of gravitational and material rigidities, i.e. $k_g/k_{\mathrm{el}}$, is termed the $Q$ enhancement factor, and for a pendulum of a single wire \cite{Saulson1990},
\begin{equation}
Q_{m}=\frac{4l}{r^2}\sqrt{\frac{mg}{E\pi}}Q_{\mathrm{mat}}
\label{dillution}
\end{equation}
where $l$ is the length of the wire, $r$ is its radius, $m$ is the mass, $E$ is the Young's modulus of the material, and $Q_{\mathrm{mat}}$ is its intrinsic quality factor of the material. It is thus evident that in order to achieve maximum dilution the choice of material, as well as minimizing (maximizing) the radius (length) of the wire have to be taken into consideration. Dissipation in the system can originate principally through energy loss from internal or external channels, and the total loss will be given by a sum of all the losses. Internal losses take into account material losses, surface losses, and thermoelastic losses. On the other hand, external losses can come from residual gas losses, clamping losses, and bonding losses. In general, the study of different loss mechanisms is critical to achieve minimum dissipation in the pendulum \cite{Cumming2012}. 


Regarding the dissipation's frequency dependence, the pendulum is known to follow the structural damping model \cite{Saulson1990, Gonzalez1994} in frequencies where higher-order modes are sufficiently sparse. Energy loss generates from internal material losses, and the dissipation is not constant (as opposed to viscous damping, where the mode is assumed to be damped by external friction) but depends on the frequency:

\begin{equation}
\gamma(\omega)=\frac{\omega^2_m}{Q_m\omega},
\label{structure}
\end{equation}
where the quality factor of the pendulum mode is related to the constant loss angle by $\phi_m=1/Q_m$. This is advantageous, since the displacement noise spectral density of a structurally damped pendulum falls faster than a viscously damped oscillator ($x_{\mathrm{th}}\propto 1/\omega^{2.5}$ vs $1/\omega^2$), lowering the noise floor of the suspension thermal noise at higher frequencies.

\begin{figure}[tb]
	\includegraphics[keepaspectratio=true,width=\columnwidth]{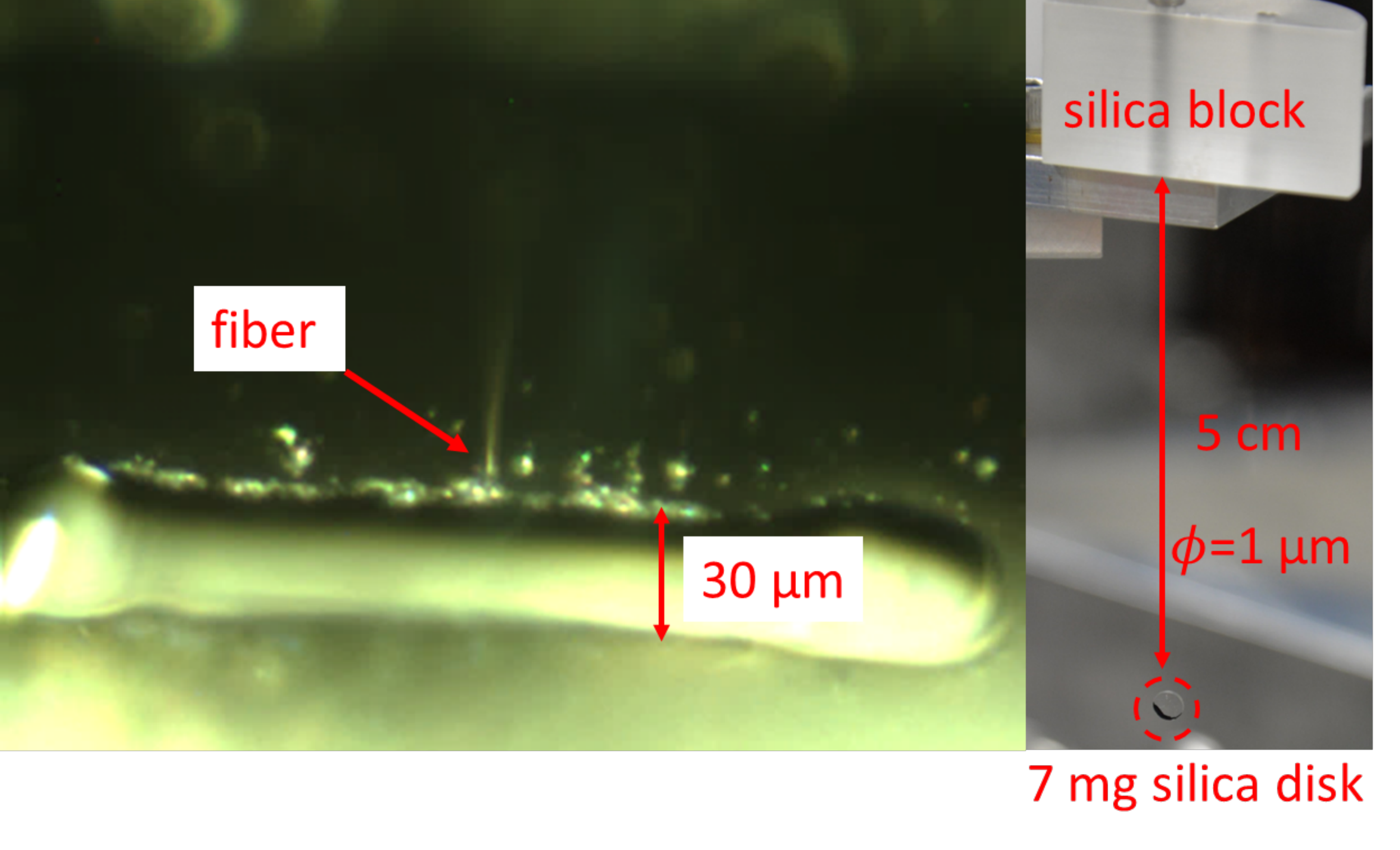}
	\caption{Picture of the welding point (left) at the test mass and the pendulum (right). The image of the welding point was taken with an optical microscope (SELMIC, SE-1300 microscope; SEL-80 objective lens).}
	\label{weld}
\end{figure}

\begin{figure*}[!t]
	\centering
	\includegraphics[width=\textwidth]{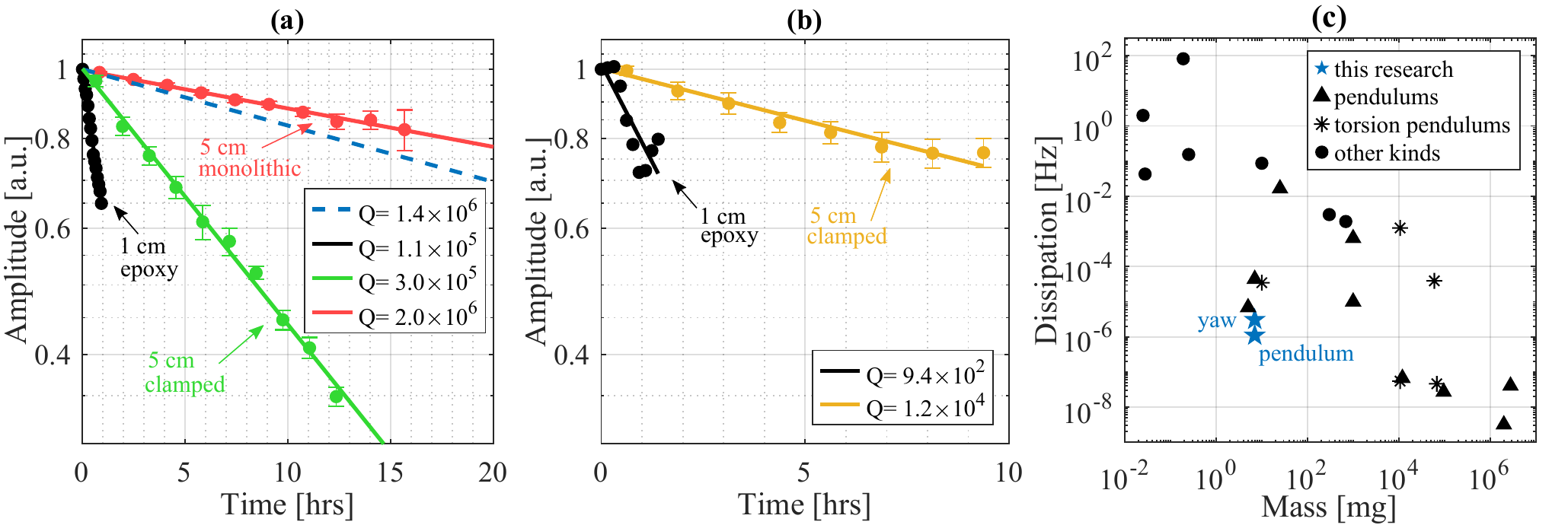}
	\caption{Logarithmic plot of ring-down measurements (a) showing a comparison between the required quality factor (blue dashed), the previously obtained value in \cite{Matsumoto2019} (black), a pendulum with the disk and fiber welded but clamped at the top (green), and the completely monolithic pendulum (red). The requirement is calculated assuming an optical spring effective frequency of 280 Hz. Plot (b) shows averaged ring-down measurements of the yaw mode of a different 5 cm long clamped pendulum (yellow) after adding the mass flow control and fabricated with the same pulling rig, versus the ring-down measurement of the previous experiment (black). Plot (c) shows the mechanical mode's dissipation $\gamma_m/2\pi$ vs mass for published experiments with macroscopic oscillators. This experiment is depicted with the star symbol. Data adapted from \cite{Imboden2014,Aspelmeyer2014,Komori2020,Braginsky1996,LeTraon2011,Agatsuma2014,Hueller2002,Rowan1997,Cagnoli2000,Newman2014,Matsumoto2019,Matsumoto2015,Altin2017,Li2018,Pontin2018,Greywall1994,Arcizet2006,Tittonen1999,Mow-Lowry2008,Corbitt2007a,Gonzalez1995,Neben2012}.}
	\label{results}
\end{figure*}

{\it Fabrication.}---
We fabricate a 1 $\mu$m fiber diameter with a length of 5 cm starting from a 125 $\mu$m diameter fused silica fiber. This is done by pulling the fiber while heating it with a hydrogen torch  (HORIBA, OPGU-7100). The fiber is pulled about 30 cm by programable motorized stages (SIGMAKOKI, SHOT-GS, OSMS26-300), and its taper region follows the model in \cite{Birks1992}. Improvements since \cite{Matsumoto2019} are the addition of a mass flow controller (HORIBA, SEC-E40MK3) to reduce surface imperfections and an increased fiber taper length of 1 cm to 5 cm. The former has improved the intrinsic material quality factor by an order of magnitude, while the latter directly affects gravitational dilution. The material quality factor, measured via a ring-down measurement of the pendulum's yaw mode, is estimated to be $ Q_{\mathrm{yaw}}=1.2 \times 10^{4}$. This value is close to the limiting $Q$ due to the fiber's surface losses $\approx 2 \times 10^{4}$ as estimated in \cite{Penn2006}. An example of this improvement is shown later in Fig. \ref{results} (b), along with measurements for other pendulums fabricated with this method. The repeatability of the fiber-pulling rig has been confirmed by SEM measurements.

Once the ultra-thin and long fiber is fabricated, we proceed to mount it on a bench implementing a $\mathrm{CO_2}$ laser (Coherent, Diamond C-30A) for welding the test mass, fiber, and the silica block support at the top. The laser spot is focused to a 30 $\mu$m beam spot, which allows localizing the welding point as shown in the left picture of Fig. \ref{weld}. The monolithic aspect of this approach is critical in reducing loss mechanisms since, in contrast with kg-scale systems \cite{Cagnoli2000,Cumming2012,Braginsky1996}, previous reports utilizing tabletop mg- to g-scale test masses have until now been unable to achieve comparable levels of dissipation \cite{Matsumoto2016,Neben2012,Nagano2016}. The test mass is a 7 mg silica disk of 3 mm in diameter and 0.5 mm in width to emulate a suspended mirror in a cavity optomechanics experiment (right of Fig. \ref{weld}).

{\it Results and discussion.}---
Fig. \ref{results} shows several averaged ring-down measurements after excitation of the pendulum. The position of the pendulum is measured by detecting the intensity modulation due to the shadow cast onto a Si photodiode (HAMAMATSU, S1223-01) from a laser (Coherent, Mephisto 500) intersecting the test mass' path. A bandpass filter is applied to the data around the resonance of interest, and the envelope of the time trace is then extracted. The results of multiple measurements are aggregated into time bins, where the average and statistical error are shown in Fig. \ref{results}, and fitted to an exponential function. The $Q$ value is then calculated from the fit parameter. Further, to neglect residual gas damping, the experiment was performed at low pressure. In our case, the experiment was performed at pressures lower than $10^{-5}$ Pa, which would limit the $Q_m$ at around $10^9 \times (\frac{\omega_m/2\pi}{2.2\mathrm{[Hz]}})$. 

Fig. \ref{results} (a) shows a comparison between the pendulum utilized in our previous report \cite{Matsumoto2019} (black data), attempts of welding only the disk and the fiber (green data) and results of the fully monolithic pendulum (red data). The former was a 1 cm long and 1 $\mu$m in diameter fused silica fiber bonded to a silica mirror by epoxy glue, and clamped at the top by a pair of stainless steel plates. That system had performed with a quality factor of $1 \times 10^5$ and had a resonance frequency of 4.4 Hz. Although the 5 cm clamped pendulum (green data) shows some amount of gain, most of the 6 fold decrease in dissipation can be explained by the increase in length of the pendulum from 1 cm to 5 cm.  The biggest gain in terms of dissipation was achieved when the top clamping parts were removed and instead welded. This agrees with the assumption that the pendulum mode has most of its bending and energy loss at the top of the fiber, not near the mass \cite{Cagnoli2000a}.  

This time, we report a 40 fold decrease in terms of dissipation, since we measure a quality factor of $Q_m=2.0 \times 10^6$ (statistical error of $\pm 4\%$), at a resonance frequency of 2.2 Hz. Fig. \ref{results} (c) shows a compilation of different representative experiments with macroscopic mechanical oscillators and their dissipation. We see that our pendulum performs with the lowest dissipation at the mg-scale. Our pendulum's parameters are close to those suggested in proposals for probing the quantum nature of Newtonian interactions by measuring gravity-induced light correlations \cite{Miao2019}. Furthermore, it surpasses the requirement of maintaining at least one coherent oscillation before thermal decoherence, since $Q_{\mathrm{eff}}\omega_{\mathrm{eff}}/2\pi=9.2\times 10^{12}$, under the same modified effective frequency of 280 Hz as in our last report \cite{Matsumoto2019}.

To calculate $Q_{\mathrm{eff}}$ and $\omega_{\mathrm{eff}}$, we work with the assumption that the effective oscillating mode is the pendulum mode as modified by the optical spring once the suspended mirror is confined in the optical trap
\cite{Matsumoto2015,Matsumoto2019,Corbitt2007,Search2009}. Because the optical spring is effectively lossless, it allows the oscillator to undergo further dilution given by the enhancement factor $ k_{\mathrm{eff}}/k_g = (\omega_{\mathrm{eff}}/\omega_m)^2$, where the effective rigidity $k_{\mathrm{eff}}=k_{\mathrm{opt}}+k_g+k_{\mathrm{el}}$, and $k_{\mathrm{opt}}$ is the optical rigidity. In the enhancement factor, we have ignored the material rigidity $k_{\mathrm{el}}$ because $k_{\mathrm{opt}}\gg k_g\gg k_{\mathrm{el}}$. Thus, the achievable quality factor scales as $Q_{\mathrm{eff}}=Q_m\times(\omega_{\mathrm{eff}}/\omega_m)^2$. 
We note here that when analyzing the pendulum mode's spectrum we observed fluctuations of its resonance frequency in the order of a few $\mu$Hz, resulting in phase decoherence. We attribute this to electrical charge up and coupling of the silica.
 However, because at our frequencies of interest the optical rigidity is much larger than the bare pendulum's rigidity ($k_{\mathrm{opt}}/k_g\approx10^4$), these fluctuations are negligible at the effective frequency even to first order, since $k_{\mathrm{eff}}=k_{\mathrm{opt}}+k_g \rightarrow \omega_{\mathrm{eff}}=\omega_{\mathrm{opt}}\sqrt{1+(\omega_m/\omega_{\mathrm{opt}})^2}$.


 In fact, although our pendulum is still two orders of magnitude away from reaching the ideal quality factor given by Eq. (\ref{dillution}) (we believe this may be due to welding losses \cite{Heptonstall2010}), the current state is enough to fulfill both requirements and further improvement would be masked by other dissipation mechanisms. Therefore, future attempts at improving thermal noise will benefit from focus on the mirror's thermal noise and the fiber's thermoelastic noise. For our system, when considering a model including nonlinear thermoelastic losses \cite{Penn2006} it is possible to tune the fiber radius to effectivly cancel out thermoelastic losses at our frequency of interest \cite{Heptonstall2014}.

\begin{figure}[th]
	\includegraphics[width=\columnwidth]{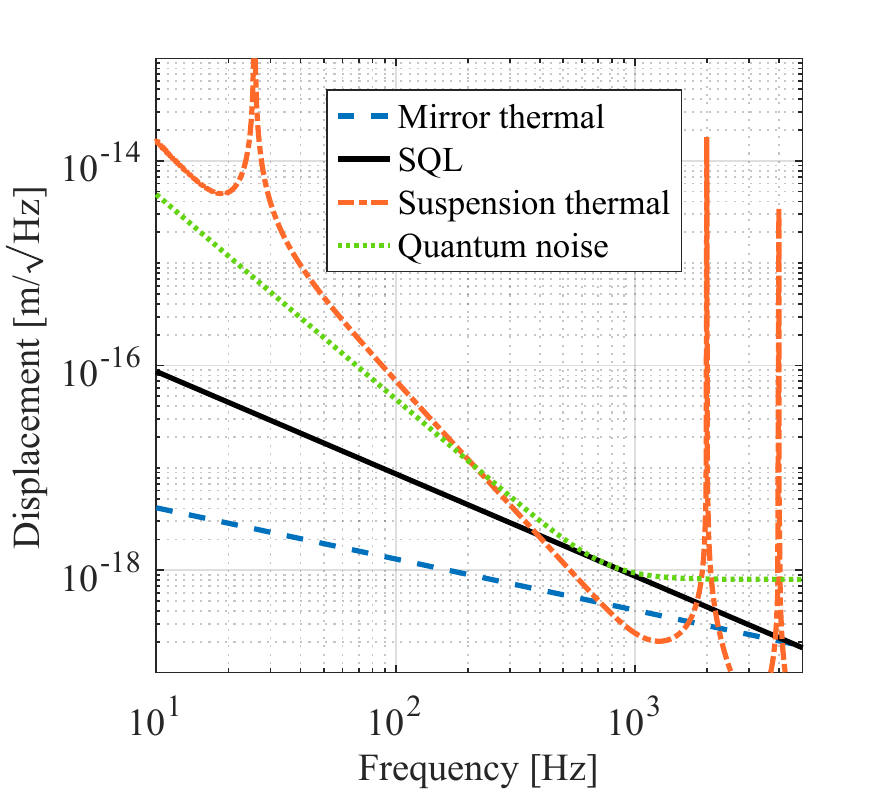}
	\caption{Noise budget for a cavity-optomechanics experiment with this pendulum as suspension for the movable mirror. Suspension thermal noise is calculated using the measured dissipation of the pendulum mode. Only the pitching mode and the first two violin modes are shown. Mirror thermal noise was calculated with a substrate loss angle of $1\times 10^{-6}$, coating loss angle of $3\times 10^{-5}$, and a beam radius of 184 $\mu$m.}
	\label{budget}
\end{figure}

In terms of dissipation, the expected value following the structural damping model (Eq. (\ref{structure})) satisfies the second requirement  (Eq. (\ref{mesurementrate})) at frequencies above 400 Hz. Fig. \ref{budget} shows a design sensitivity considering higher-order modes (i.e. pitching mode and violin modes), mirror thermal noise (substrate plus coating thermal noise \cite{Numata2003}), and quantum noise (limited at low frequencies by quantum radiation pressure noise and at high frequencies by shot noise). We note that to achieve the design level of mirror thermal noise, state of the art coatings like crystalline coatings \cite{Cole2013,Chalermsongsak2016} should be implemented. Suspension thermal noise is calculated using the analytic model in \cite{Gonzalez1994}, derived by solving the elastic beam equation with boundary conditions corresponding to a rigid mass of finite size. Due to the 40 fold decrease in dissipation, this pendulum's suspension thermal noise is estimated to be roughly 6 times lower than that of our previous report \cite{Matsumoto2019}. Also, the improvement in the material quality factor suggests this fabrication method can be advantageous for testing CSL models with mg-scale torsion pendulums \cite{Komori2020}. In Fig. \ref{budget} we do not include the optical rigidity, and see that our pendulum's thermal noise goes below the SQL between 400 - 1800 Hz, meaning quantum fluctuations dominate the noise spectrum. Because the  optical spring only changes the mechanical susceptibility, quantum-limited sensing can be achieved around 1 KHz on an optically trapped pendulum's resonance. 

For quantum-control experiments, an on-resonance probe beam of 0.2 mW (shot noise limit of about $ 4\times10^{-8}/\sqrt\mathrm{Hz}$ in relative intensity units) gives the desired level of quantum noise for a critically coupled cavity with finesse of 5000. Necessary frequency noise must be below $\sim6$ $\mathrm{mHz}/\sqrt\mathrm{Hz}$ around the frequency band of interest for a target design cavity round-trip length of 10 cm. This is feasible with traditional intensity stabilization techniques \cite{Kwee2009}, and frequency stabilization utilizing rigid cavities \cite{Numata2003}. An additional beam with detuning $6\kappa$, where $\kappa$ is the cavity amplitude decay rate, and input power  of $100$ mW can be used to create an optical spring around 750 Hz, where the large detuning is to suppress the trapping beam's back-action \cite{Braginsky1999,Chang2012}. The probe beam's signal can be read with homodyne detection, then high-pass filtered to create a force proportional to the oscillator's velocity, and fed-back to an actuator. Since Eq. (\ref{mesurementrate}) is satisfied, ground-state cooling of the confined mode is achievable. 

Lastly, other applications include implementation of proposals that utilize the free-mass region of the oscillator. For example,  because the thermal noise is less than the SQL, the optomechanical system can be used as testbed for quantum nondemolition measurements  \cite{Braginsky1980,Kimble2001} at several hundred Hz (frequency band of interest for GW detectors). Similarly, implementing two pendulums like this one as end mirrors in a power-recycled Fabry-Perot-Michaelson configuration, entanglement between the differential and common modes of macroscopic test masses, as proposed in \cite{Muller2008}, will also be possible.



{\it Conclusion.}---
We report the fabrication of a completely monolithic mg-scale pendulum meeting requirements for performing quantum control experiments. To the best of our knowledge, this is the lowest dissipation ever achieved and the highest $Q$ at room temperature for a mechanical oscillator of this mass scale. Combined with the optical spring effect, it can open the door for experimentation in the intersection between quantum theory and gravity. 

{\it Acknowledgments.}---
We thank Takao Aoki, Yuuji Matsuura, Yuanyuan Guo, and Hiroki Takahashi for stimulating discussions about fiber fabrication and laser welding. 
We also thank Masakazu Sugawara, and Tenma Kanai for help with the fiber fabrication. This research was supported by  JSPS KAKENHI Grant Number 19H00671, PRESTO JST, and JST CREST Grant Number JPMJCR1873. 

\bibliography{biblio}

\end{document}